# On-the-fly Query-Based Debugging with Examples

Raimondas Lencevicius
Nokia Research Center, 5 Wayside Road, Burlington, MA 01803
Raimondas.Lencevicius@nokia.com
http://alumni.cs.ucsb.edu/~raimisl

**Abstract.** Program errors are hard to find because of the cause-effect gap between the time when an error occurs and the time when the error becomes apparent to the programmer. Although debugging techniques such as conditional and data breakpoints help to find error causes in simple cases, they fail to effectively bridge the cause-effect gap in many situations. Query-based debuggers offer programmers an effective tool that provides instant error alert by continuously checking inter-object relationships while the debugged program is running. To enable the query-based debugger in the middle of program execution in a portable way, we propose efficient Java class file instrumentation and discuss alternative techniques. Although the on-the-fly debugger has a higher overhead than a dynamic query-based debugger, it offers additional interactive power and flexibility while maintaining complete portability. To speed up dynamic query evaluation, our debugger implemented in portable Java uses a combination of program instrumentation, load-time code generation, query optimization, and incremental reevaluation. This paper discusses on-the-fly debugging and demonstrates the query-based debugger application for debugging Java gas tank applet as well as SPECjvm98 suite applications.

## 1  Introduction

Many program errors are hard to find because of a cause-effect gap between the time when the error occurs and the time when it becomes apparent to the programmer by terminating the program or by producing incorrect results [Eis97]. The situation is further complicated in modern object-oriented systems which use large class libraries and create complicated pointer-linked data structures. If one of these references is incorrect and violates an abstract relationship between objects, the resulting error may remain undiscovered until much later in the program's execution.

Consider trying to debug the javac Java compiler, a part of Sun's JDK distribution. During a compilation, this compiler builds an abstract syntax tree (AST) of the compiled program. Assume that this AST is corrupted by an operation that assigns the same expression node to the field right of two different parent nodes (Figure 1). The parent nodes may be instances of any subclass of BinaryExpression; for example, the parent may be an AssignAddExpression object or a DivideExpression object, while the child could be an IdentifierExpression. The compiler traverses the AST many times, performing type checks and inlining transformations. During these traversals, the child expression will receive contradictory information from its two parents. These contradictions may eventually become apparent as the compiler indicates errors in correct Java programs or when it generates incorrect code. But even after discovering the existence of the error, the programmer still has to determine which part of the program originally caused the problem. How can we help programmers to find such errors as soon as they occur?

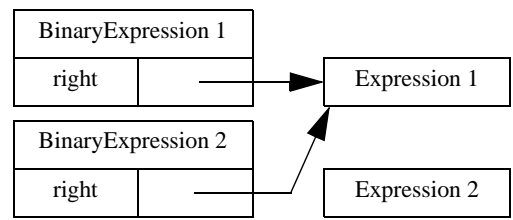

**Figure 1.**  Possible error in javac AST

We proposed [LHS97][LHS99] query-based debugging which can overcome these problems. Similar to an SQL database query tool, a query-based debugger (QBD) finds all object tuples satisfying a given boolean constraint expression. For example, the query

    BinaryExpression* e1, e2.    e1.right == e2.right && e1 != e2

would find the objects that are erroneously shared by two parents through the right field. A dynamic query-based debugger continually updates the results of queries as the program runs, and can stop the program as soon as the query result changes. To provide this functionality, the debugger finds all places where the debugged program changes a field that could affect the result of the query and uses sophisticated algorithms to incrementally reevaluate the query. Therefore, a dynamic query-based debugger finds the javac AST bug as soon as the faulty assignment occurs, and it does so with minimal programmer effort and low program execution overhead. The on-the-fly debugger adds a capability to stop the javac program just before the AST construction phase and enable the query. It also allows to change the query later.

We have implemented such a dynamic query-based debugger for Java. Our prototype is portable—written in 100% pure Java with no JVM modifications, and surprisingly efficient. Experiments with large programs from the SPECjvm98 suite [SPEC98] show that selection queries are very efficient for most programs, with a slowdown of less than a factor of two in most experiments. More complicated join queries are less efficient but still practical for small query domains or programs with infrequent queried field updates. These results are further presented in [LHS97][LHS99].

However, the implementation of the dynamic query-based debugger required users to specify queries before the program execution starts. Queries were enabled from the beginning of the program execution and remained active until its end. These requirements diminished the usefulness of the debugger because users could not restrict queries to parts of the program execution and could not ask new queries in the middle of a program run. In a long-running program, or in a hard-to-reproduce test case, the ability to add queries on the fly would save substantial amount of debugging time. Conventional debuggers allow programmers to place simple breakpoints and check variable values during program execution but they do not support complex queries.

An on-the-fly debugger implementation proposed in this paper makes the debugger fully interactive while maintaining debugging portability for different Java virtual machines and operating systems. With the on-the-fly debugger, programmers can stop a program at any time using conventional breakpoints or another already enabled query, enter a query or change it later when more information about the error becomes available. The system is portable without any changes across Java virtual machines, operating systems, and CPUs. The system was tested on Sun SPARC system running Solaris with Solaris Java 1.2 VM and Intel system running Windows NT with Sun Java 1.2 VM.

The on-the-fly debugging overhead measured for large programs from the SPECjvm98 suite [SPEC98] show that the instrumented programs with inactive debugger suffer a median overhead of 25% (less than 70% for all applications). The selection query overheads range up to factor 9.5 with a median of 5.5. More complicated join queries are less efficient but may be practical for small query domains or short debugging runs.

In this paper, we discuss the on-the-fly debugger, its implementation and efficiency. We also demonstrate the query-based debugger for Java on a gas tank applet as well as SPECjvm98 applications.

## 2  Query Model

Query-based debugging uses the query model proposed in [LHS97]. The query syntax is as follows:

    <Query>                    ::==    <DomainDeclaration> **{ ; **<DomainDeclaration> **} .**
                                              <ConditionalExpression>
    <DomainDeclaration>    ::==    <ClassName> [*] <DomainVariableName>
                                              { <DomainVariableName> }

The query has two parts: one or more DomainDeclarations that declare variables of class ClassName, and a ConditionalExpression. The first part is called the *domain part* and the second the *constraint part*. Consider another javac query:

> FieldExpression fe; FieldDefinition fd.
> fe.id == fd.name && fe.type == fd.type && fe.field != fd

The first part of the query defines the *search domain* of the query, using universal quantification. The domain part of the above example should be read as "for all FieldExpressions fe and all FieldDefinitions fd...". FieldExpression is a class name and its domain contains all instances of the class. If a "*" symbol in a domain declaration follows the class name (as in the javac query discussed in the introduction), the domain includes all objects of subclasses of the domain class, otherwise the domain contains only objects of the indicated class itself. The second part of the query specifies the constraint expression to be evaluated for each tuple of the search domain. Constraints are arbitrary Java conditional expressions as defined in the Java specification §15.24 [GJS96] with certain syntactic restrictions (see [LHS99]).

Semantically, the expression will be evaluated for each tuple in the Cartesian product of the query's individual domains, and the query result will include all tuples for which the expression evaluates to true (similarly to an SQL select query). Conceptually, the debugger reevaluates a query after the execution of every bytecode, ensuring that no result changes are unnoticed. The debugger stops the program whenever the result changes. In practice, the debugger reevaluates the query as infrequently as possible without violating these semantics [LHS99].

We refer to queries with a single domain variable as *selection queries*; following common database terminology, we call the rest of the queries *join queries* because they involve a join (Cartesian product with selection) of two or more domain variables. Join queries with equality constraints only (e.g., p1.x == p2.x) are called *hash joins*. They can be evaluated more efficiently using a hash table [LHS97].

## 3 Examples

In this section, we present examples of debugging situations handled by dynamic query debuggers.

### 3.1 Javac Compiler

What are examples of inter-object constraint violations that may be difficult to trace back to their origins? We have already discussed one possible error in the javac Java compiler in the introduction. Another error that could occur in javac involves the relationship between FieldExpression and FieldDefinition objects. Consider a situation where a FieldExpression object no longer refers to the FieldDefinition object that it should reference. Due to an error, the program may create two FieldDefinition objects such that the FieldExpression object refers to one of them, while other program objects reference the other FieldDefinition object (Figure 2). In other words, javac maintains a constraint that a FieldExpression object that shares the type and the identifier name with a FieldDefinition object must reference the latter through the field field. We can detect a violation of this constraint using the following query:

> FieldExpression fe; FieldDefinition fd.
> fe.id == fd.name && fe.type == fd.type && fe.field != fd

This complicated constraint can be specified and checked with a simple dynamic query.

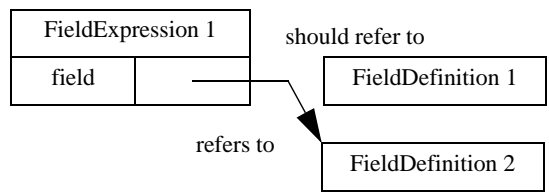

**Figure 2.** Another possible error in javac AST

## 3.2 Ideal Gas Tank Example

Another program we examined is an applet simulating a tank with ideal gas molecules. (Figure 3). Although this applet is a simple simulation of gas molecules moving in the tank and colliding with the tank walls and each other, it has some interesting inter-object constraints. First, all molecules have to remain within the tank, a constraint that can be specified by a simple selection query:

    Molecule* m.    m.x < 0 || m.x > X_RANGE || m.y < 0 || m.y > Y_RANGE

Another constraint requires that molecules do not occupy the same position as other molecules. Even this simple application may violate the constraint in different places: in the regular molecule move method, in a method that handles molecule bounces from the walls, and so on. The following query discovers the constraint violation:

    Molecule* m1 m2.    m1.x == m2.x && m1.y == m2.y && m1 != m2

This constraint is interesting because its violation is a transient failure. Transient failures disappear after some period of time, so even though the program behaves differently than the programmer expected, queries will not be able to detect failures if they are asked too late. The molecule collision error is such a transient failure—it will disappear as the molecules continue to move. However, the applet will behave erroneously: for example, molecules that should have collided with each other will pass through each other. Dynamic queries are necessary to find transient failures, as delayed query reevaluation may fail to detect the error entirely.

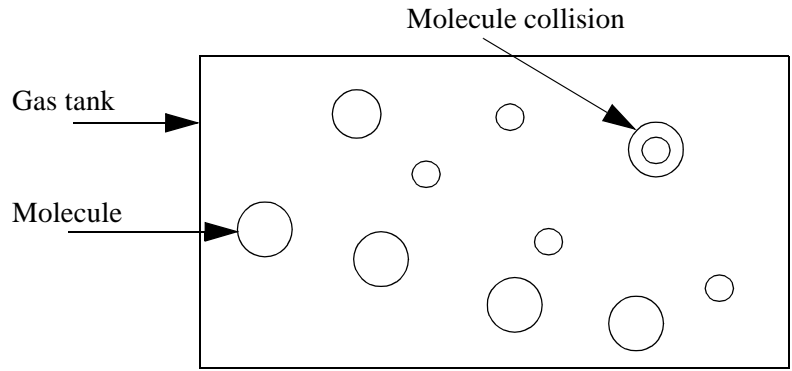

**Figure 3.** Error in molecule simulation

## 4 Implementation

In this section, we summarize the implementation of the dynamic query-based debugger and present the changes required for the on-the-fly debugging. We have implemented query-based debuggers in pure Java. We used Java's ability to write custom class loaders [LB98] to perform load-time code instrumentation. Java's bytecode class files proved simple to instrument. The debugger creates custom query evaluation code by using load-time code generation. The prototype implementation only demonstrates the query-based debugging and on-the-fly capability. For production use, the system should be integrated into a full-fledged debugging suite.

### 4.1 General Structure of the System

Figure 4 shows a data-flow diagram of the on-the-fly debugger. To debug a program, the user runs a standard Java virtual machine with a custom class loader. The custom class loader loads the user program and instruments the bytecodes loaded by adding debugger invocations for each object creation and field assignment. The class loader also generates and compiles custom debugger code. After loading, the Java virtual machine executes the instrumented user program. Whenever the program reaches instrumentation points, it checks whether the debugger is active and if so, invokes the custom debugger code, which calls other debugger runtime libraries to reevaluate the query and to generate query results.

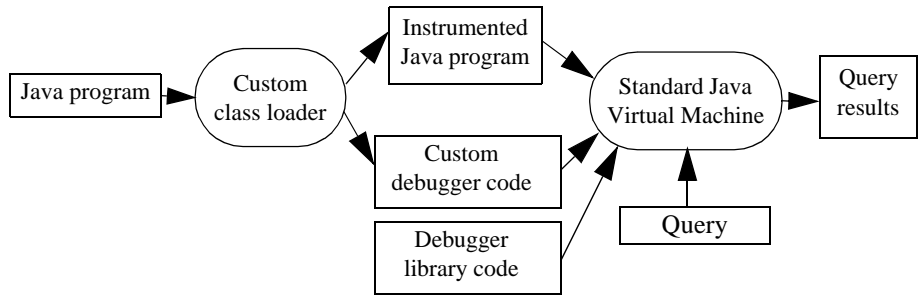

**Figure 4.** Data-flow diagram of on-the-fly debugger

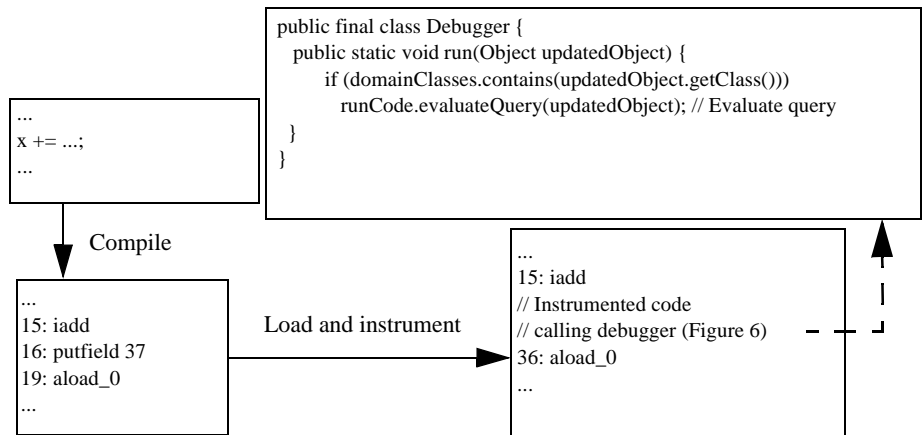

**Figure 5.** Java program instrumentation

### 4.2 Java Program Instrumentation

The original debugger implementation could not support on-the-fly debugging because the debugger had to know a query and its change set (section 4.3) to instrument class files at load time. The class loader then instrumented the assignments to the monitored fields and the creations of the domain objects while loading Java class files. Class files cannot be instrumented after loading without changing the Java virtual machine. On the other hand, changing the JVM would compromise the portability of the debugger across different virtual machines.

If queries are unknown before the program execution, the debugger has to invoke the debugger after all events that may change the result of any query. The on-the-fly debugger instruments class files by inserting debugger invocations after each assignment to an object field. The system also inserts debugger invocations after each call to a constructor of an object. Figure 5 shows an example of such instrumentation process for a Java method. To instrument class files, the loader transforms them in memory into a malleable format using modified class-file handling tools from Binary Component Adaptation library [KH98]. Then the loader finds all putfield bytecodes and adds invokestatic bytecodes invoking debugger code after the putfield bytecodes. Figure 6 shows the exact bytecodes that replace a single putfield. The system also inserts such debugger invocations after each call to a constructor. The debugger adds a reference to the run method of the Debugger class into the constant pool of the instrumented class. The method takes as an argument the object updated by the putfield—a Molecule object in the example. This object is on the stack before execution of the putfield, so a copy of it can be obtained by stack manipulation (Figure 6) that duplicates the top two values on the stack and then discards the top one (the value assigned by the putfield). The debugger determines the correct types

```
16: getstatic 133 // Get debugger activation flag
19: ifeq 14 (33) // If debugger disabled go to bytecode 33
// If debugger enabled get debugger parameters,
// perform putfield, and invoke debugger
22: dup2 // Duplicate modified object and assigned value
          on the stack
23: putfield 35 // Perform original putfield
26: pop // Pop the assigned value
27: invokestatic 127 // Invoke debugger
30: goto 6 (36) // Go to bytecode 36
33: putfield 35 // Perform original putfield
36: // end of instrumented block
```

**Figure 6.** On-the-fly debugging instrumentation

of objects from the class file's constant pool. After instrumentation, the class loader transforms the code back into the class file format and passes the image to the default defineClass method.

To limit the overhead if the on-the-fly debugger is not enabled, the instrumentation inserts a test around each putfield code. If the debugger is not enabled, the program executes only two additional bytecodes per each putfield bytecode: a load of a debugger flag (getstatic) and a conditional jump (ifeq) to the original putfield. Figure 6 shows the instrumentation performed on a single putfield bytecode. The "fast path" has only two extra bytecodes. However, if the debugger is enabled, the overhead is higher. In this case, the debugger has to replicate the reference to the updated object, pass it to the debugger's run method and invoke that method. Then the debugger determines if modified object belongs to one of the query's domain classes, and if so proceeds with further evaluation.

The next section describes how the debugger determines which assignments and constructors influence the query result and when the query is reevaluated.

### 4.3 Change Monitoring and Query Evaluation

The debugger updates the query result every time the debugged program performs an operation that may affect the query result. Thus, the program being debugged has to invoke the debugger after every event that could change the query result. The query result may change because some object assigns a new value to one of its fields or because a new object is constructed. However, not all field assignments and object creations affect the query. We call the set of constructors and object field assignments affecting the results of a query the query's *change set*. Although the on-the-fly debugger instruments all assignments and all constructors as a conservative change set for all queries, we are interested in a minimal change set for query evaluation. Such a change set contains only constructors of domain objects and assignments to domain object fields referenced in a query[*].

Consider the Molecule query:

Molecule* m1 m2.    m1.x == m2.x && m1.y == m2.y && m1 != m2

The change set of this query consists of the constructors of the Molecule class and its subclasses as well as assignments to Molecule fields x and y. Assignments to other molecule fields such as color do not belong to the change set.

---

[*] More complicated change sets are discussed in [LHS99]

The change set of a query tells the debugger what assignments and constructors are relevant for query evaluation. Before definition of any queries, the on-the-fly debugger does not execute evaluation code at all. When a query is defined, the debugger uses the minimal change set to decide when a query should be reevaluated.

To support on-the-fly debugging, the debugger keeps collections of objects belonging to all classes. These collections are necessary to evaluate queries. Since the standard Java debugging API does not allow debuggers to retrieve all objects of a class, debuggers have to track creation of all program objects to have access to all query domain objects. Every time an object is created, the program invokes the debugger which places the new object into collection according to its class. During query evaluation a debugger uses object collections to iterate through all domain objects. To maintain query correctness and to facilitate garbage collection, the debugger allows the garbage collector to delete dead objects from domain collections. Object tracking, although inexpensive by itself, becomes costly because of the excessive memory use—for each object created by a program, the debugger has to maintain a WeakReference object and space in a domain collection. Referring to domain objects through weak references allows the Java virtual machine garbage collector to collect all objects that are referenced only by the debugger. However, even though domain objects are garbage collected, the weak references themselves remain in the collection, so the collection grows as the program runs. Some programs like the gas tank simulation (section 3.2) create so many temporary objects that weak references fill all available memory. To prevent such internal garbage, a more sophisticated implementation uses an internal "garbage collector" to recycle the weak references no longer pointing to the reachable objects. Unfortunately, the internal garbage collection of weak references adds an additional speed overhead and should be used only when the program runs out of memory without it.

The query execution in the on-the-fly debugger is the same as in the dynamic debugger [LHS99]. It becomes more important to quickly check whether an object that caused a debugger invocation belongs to a relevant domain. If the object does not belong to the query domain, the debugger immediately returns to the execution of the user program without reevaluating the query.

To summarize, we use the change set of the query to reevaluate the query after interesting events. The instrumented program calls the debugger after every event that could change the result of the query, and the debugger reevaluates the query during each call if the change affects query domains.

### 4.4 Alternative Implementations

On-the-fly debugging could be implemented using alternative techniques that may increase the debugger efficiency. One approach would be to change the Java virtual machine. Even though we did not pursue this approach because of its lack of portability, JVM changes may lead to the most efficient implementations. These changes could be simple or sophisticated. A simple JVM change would allow the debugger to retrieve all objects of a class[*]. Such capability would remove the necessity to track all objects of all classes and would reduce both the direct object tracking overhead and excessive memory use by weak references. More sophisticated JVM changes would allow to instrument already loaded classes and avoid the overhead of extra bytecodes surrounding each putfield bytecode.

JVM changes are not portable. An alternative technique to speed up a debugger would be to use *shadow* classes. In other words, while the debugger is not enabled, the program would execute the code that is instrumented to check the debugger activation only at the beginning of the methods and possibly at the back branches of the loops. When the debugger is enabled, it would generate fully instrumented versions of the classes. Such fully instrumented shadow methods would be invoked through the redirection at the beginning of the regular methods. This method reduces the overhead of the instrumented putfield execution but does not solve the problem of object tracking. Also, the debugger activation would be delayed until the instrumentation point is reached. Due to this delay, a debugger could miss some errors.

---

[*] We implemented it for JDK 1.1.5 during the initial design of a query-based debugger. Such collection may contain dead objects, but so can current domain collections.

# 5 Experimental Results

In our experiments, we used the same queries as in [LHS99] drawn from SPECjvm98 suite and other applications. The selected queries in complexity and overhead cover the range of queries asked in debugging situations. The query set contains selection queries with low and high cost constraints. The test also includes hash-join and nested-join queries with different domain sizes. The queries check programs that range from small applets to large applications and (for stress-tests) microbenchmarks. These applications invoke the debugger with frequencies ranging from low to very high, where a query has to be evaluated at every iteration of a tight loop. Consequently, the experimental results obtained for the test set should indicate the range of performance to be expected in real debugging situations.

For our tests we used an otherwise idle Sun Ultra 2/2300 machine (with two 300 MHz UltraSPARC II processors) running Solaris 2.6 and Solaris Java 1.2 with JIT compiler (Solaris VM (build Solaris_JDK_1.2_01, native threads, sunwjit)) [Sun99]. Execution times are elapsed times and were measured with millisecond accuracy using the System.currentTimeMillis() method.

## 5.1 Benchmark Queries

To test the on-the-fly query-based debugger, we selected a number of structurally different queries (Table 1) for a number of different programs (Table 2):

- Queries 1 and 13 check a small ideal gas tank simulation applet that spends most of the time calculating molecule positions and assigns object fields very infrequently. It has 100 molecules divided among Molecule1, Molecule2 and Molecule3 classes. The application performs 8,000 simulation steps.
- Queries 2 and 14 check the Decaf Java subset compiler, a medium size program developed for a compiler course at UCSB. The Token domain contains up to 120,000 objects.
- Query 3 checks the Jess expert system, program from the SPECjvm98 suite [SPEC98].
- Queries 4–10, and 16–17 check the compress program from the SPECjvm98 suite. Our queries reference frequently updated fields of compress.
- Queries 11–12 and 15 check the ray tracing program from the SPECjvm98 suite. The Point domain contains up to 85,000 objects; the IntersectPt domain has up to 8,000 objects.
- Queries 18–20 check artificial microbenchmarks. These microbenchmarks stress test debugger performance by executing tight loops that continuously update object fields.

Structurally, queries can be divided into the following classes:

- Queries 1–12 and 18 are simple one-constraint selection queries with a wide range of constraint complexities. Queries 4–7 have increasing constraint cost while referencing the same field. For example, query 5 is more costly than query 4 because it invokes a method to compare an object field to an integer. The most costly constraint in query 7 performs expensive mathematical operations before performing a comparison. Queries 8 and 9 have very similar constraints, but differ 4.8 times in debugger execution frequency. Here, by "debugger execution frequency" we mean the frequency of events in the original program that would trigger a debugger invocation and then perform the query evaluation, i.e., the frequency of assignments to relevant fields with no debugger overhead. Query 12 compares the parameter of the method to the distance of a point to the origin. This query combines costly mathematical operations with increased debugger execution frequency, because its result depends on all three coordinates of Point objects.
- Queries 13–17 and 19–20 are join queries. Queries 13–16 and 19 can be evaluated using hash joins. The evaluation of queries 17 and 20 has to use nested-loop joins. For join queries, the slowdown depends both on the debugger invocation frequency and sizes of the domains. Queries 13–14 have low invocation frequencies; queries 15–17, 19–20 have high invocation frequencies. Queries 14 and 15 have large domains.

## 5.2 Query Overhead

To evaluate the on-the-fly debugger, we performed the following measurements. First of all, since programs instrumented by the debugger suffer a slowdown even when the debugger is not enabled, we measured this slowdown. Table 2 shows slowdowns together with the total field assignment frequencies for SPECjvm98

| Query | On-the-fly Slowdown | DQBD Slowdown | Invocation frequency (events / s) |
|---|---|---|---|
| 1. Molecule1 z.    z.x > 350 | 3.23 | 1.02 | 15,000 |
| 2. Id x.    x.type < 0 | 1.83 | 1.11 | 16,000 |
| 3. spec.benchmarks._202_jess.jess.Token z.    z.sortcode == -1 | 4.05 | 1.25 | 169,000 |
| 4. spec.benchmarks._201_compress.Output_Buffer z.    z.OutCnt < 0 | 6.3 | 1.18 | 1,900,000 |
| 5. spec.benchmarks._201_compress.Output_Buffer z.    z.count() < 0 | 5.48 | 1.27 | |
| 6. spec.benchmarks._201_compress.Output_Buffer z.    z.lessOutCnt(0) | 5.72 | 1.37 | |
| 7. spec.benchmarks._201_compress.Output_Buffer z.    z.complexMathOutCnt(0) | 9.36 | 5.83 | |
| 8. spec.benchmarks._201_compress.Compressor z.    z.in_count < 0 | 5.58 | 1.18 | 933,000 |
| 9. spec.benchmarks._201_compress.Compressor z.    z.out_count < 0 | 5.54 | 1.10 | 196,000 |
| 10. spec.benchmarks._201_compress.Compressor z.    z.complexMathOutCount(0) | 9.54 | 1.83 | |
| 11. spec.benchmarks._205_raytrace.Point p.    p.x == 1 | 4.82 | 1.23 | 787,000 |
| 12. spec.benchmarks._205_raytrace.Point p.    p.farther(100000000) | 4.82 | 1.98 | 2,300,000 |
| 13. Molecule1 z; Molecule2 z1.    z.x == z1.x && z.y == z1.y && z.dir == z1.dir && z.radius == z1.radius    (33x33 hash join) | 21.82 | 2.13 | 54,000 |
| 14. Lexer l; Token t.    l.token == t && t.type == 27    (120,000x600 hash join) | 6.4 | 3.43 | 25,000 |
| 15. spec.benchmarks._205_raytrace.Point p; spec.benchmarks._205_raytrace.IntersectPt ip.    p.z == ip.t && p.z < 0    (85,000x8,000 hash join) | Inf | 229 | 350,000 |
| 16. spec.benchmarks._201_compress.Input_Buffer z; spec.benchmarks._201_compress.Output_Buffer z1.    z1.OutCnt == z.InCnt && z1.OutCnt < 100 && z.InCnt > 0    (1x1 hash join) | 384 | 157 | 1,500,000 |
| 17. spec.benchmarks._201_compress.Compressor z; spec.benchmarks._201_compress.Output_Buffer z1.    z1.OutCnt < 100 && z.out_count > 1 && z1.OutCnt / 10 > z.out_count    (1x1 join) | 263 | 77 | 2,600,000 |
| 18. Test5 z.    z.x < 0 | 28 | 6.4 | 42,000,000 |

**Table 1.** On-the-fly query overhead

This table gives program slowdown for different queries. First column gives the query, second column - the program slowdown with this query enabled in the on-the-fly debugger, third column - the slowdown for the same query in dynamic query-based debugger, fourth column - debugger invocation frequency if there was no overhead. For example, query 3 has on-the-fly slowdown of 4.05 times, dynamic query slowdown of 25% and invocation frequency of 169000 times per second.

| Query | On-the-fly Slowdown | DQBD Slowdown | Invocation frequency (events / s) |
|---|---|---|---|
| 19. TestHash5 th; TestHash1 th1.    th.i == th1.i    (1x20 hash join) | 935 | 228 | 40,000,000 |
| 20. TestHash5 th; TestHash1 th1.    th.i < th1.i    (1x20 join) | 935 | 930 | |

**Table 1.** On-the-fly query overhead

This table gives program slowdown for different queries. First column gives the query, second column - the program slowdown with this query enabled in the on-the-fly debugger, third column - the slowdown for the same query in dynamic query-based debugger, fourth column - debugger invocation frequency if there was no overhead. For example, query 3 has on-the-fly slowdown of 4.05 times, dynamic query slowdown of 25% and invocation frequency of 169000 times per second.

| Application | Size (Kbytes) | Total number of field assignments | Total assignment frequency (field assignments per second) | Original program execution time (s) | Disabled debugger slowdown | Enabled debugger slowdown |
|---|---|---|---|---|---|---|
| 1. Compress | 17.4 | 392,000,000 | 7,800,000 | 50.4 | 1.70 | 3.14 |
| 2. Jess | 387.2 | 25,000,000 | 1,100,000 | 22.45 | 1.30 | 1.54 |
| 3. Db | 12 | 67,000 | 897 | 72 | 1.0 | 1.0 |
| 4. Javac | 548 | 100,000,000 | 2,600,000 | 38 | 1.27 | 1.62 |
| 5. Mpegaudio | 117 | 148,000,000 | 2,600,000 | 49.5 | 1.25 | 1.96 |
| 6. Jack | 127 | 5,700,000 | 214,000 | 26 | 1.15 | 1.19 |
| 7. Ray tracer | 55.7 | 44,000,000 | 2,200,000 | 17 | 1.12 | 1.62 |
| 8. Decaf | 55 | 7,900,000 | 528,000 | 15 | 1.15 | 1.40 |
| 9. Ideal gas tank | 14.3 | 4,000,000 | 70,000 | 57 | 1.27 | 2.0 |
| 10. Microbenchmark | 1 | 100,000,000 | 40,000,000 | 2.4 | 3.28 | 11.14 |

**Table 2.** On-the-fly debugging overhead

Column one gives application, column two - size of its class files, column three - total number of field assignments during the program's execution, column four - field assignment frequency, column five - original program execution time, column six - the slowdown with instrumentation but without debugger invocations, column seven - slowdown with debugger invocations, but with no query evaluations. For example, compress has the size of 17.4 Kilobytes, has 392 million field assignments performed 7.8 million times per second. The program executes in 50.4 seconds. Instrumented compress has a slowdown of 70% and with debugger enabled 3.14 times.

programs as well as microbenchmarks. This table indicates that adding two bytecodes (getstatic-ifeq) before each putfield costs less than 70% for applications with a median overhead[*] of 25% and 3.3 times for a microbenchmark[†].

---

[*] Although the meaning of "*median*" numbers can be interpreted differently, we provide median overhead numbers in this paper, because they seem to give a general indication of system performance.

[†] In the current JVM/JIT, the insertion of the same two bytecodes *after* the putfield bytecode instead of *before* it reduces overhead from 70% to 40% for compress. This phenomenon does not occur with the JIT disabled, and we cannot explain it.

If the debugger is enabled, but the query is never evaluated, for example, because domains contain only non-instantiated classes, programs have a larger slowdown. In this case, the instrumented byte code invokes the debugger debug method. This method at the very least checks whether the changed object is a domain object. With the debugger enabled, but no query ever evaluated, the applications have a slowdown ranging up to 3.14 with a median overhead of 62%. The microbenchmark slowdown is 11.14, a number increased by the fact that the microbenchmark assigns to a long integer field that needs more complicated instrumentation and has higher slowdown after it. Both experiments above do not include the object tracking overhead.

Finally, if a query needs to be reevaluated, the additional slowdown to reevaluate the query depends on the query. A large part of the query reevaluation time is consumed by the domain collection maintenance and by extra garbage collection. For example, in selection query 11, 36% of the query evaluation time was due to the object collection and additional GC overhead, 17% of the time was consumed by the domain class check. Overheads for all queries are given in Table 1. Selection overhead ranges up to factor 9.5 with a median of 5.5. Selection query overheads almost totally depend on the program executed and neither on the query itself, nor on the query reevaluation frequency. The low cost of selection reevaluation seems to be overshadowed by the overheads of on-the-fly instrumentation, domain collection maintenance, and garbage collection. These areas could yield substantial optimization benefits.

Join query overheads are very high. Query 15 was aborted after running for more than a day. However, on-the-fly debugging may be used when programmers only need to check query results during a part of program execution. We compared the overheads of the on-the-fly debugger with the dynamic query debugger which instruments only the relevant field assignments but has to run for the whole program execution time. The on-the-fly debugger overhead is on average four times higher than the overhead of the dynamic debugger. The performance is much closer for expensive queries. If programmer wants to inspect only small part of the program runtime, the on-the-fly method may be more useful than the dynamic debugging.

## 6 Related Work

We are unaware of other work that directly corresponds to dynamic query-based debugging and its on-the-fly extension. The query-based debugging model and its non-dynamic implementation are presented in a previous paper [LHS97]. The dynamic query based debugging is presented in [LHS99].

Extensions of object-oriented languages with rules as in R++ [LMP97] provide a framework that allows users to execute code when a given condition is true. However, R++ rules can only reference objects reachable from the root object, so R++ would not help to find the javac error we discussed. Due to restrictions on objects in the rule, R++ also does not handle join queries.

Sefika et al. [SSC96] implemented a system allowing limited, unoptimized selection queries about high-level objects in the Choices operating system. The system dynamically shows program state and run-time statistics at various levels of abstraction. Unlike our dynamic query-based debugger, the tool uses instrumentation specific to the application (Choices). Sefika's system allows on-the-fly queries because the underlying system is instrumented for information gathering.

On-the-fly debugging idea is based on the design of the commercial debuggers (e.g., gdb) that allow programmers to stop the program and to add breakpoints before further execution [Kes90] and to check the values of different variables at a breakpoint. Such capabilities are also available in data breakpoint debuggers [WLG93]. Dynamic query-based debugging extends work on data breakpoints [WLG93]—breakpoints that stop a program whenever an object field is assigned a certain value. Debuggers that instrument source code programs have also been proposed [FB63]. However, the problem of allowing a portable implementation of on-the-fly debugging through automatic instrumentation of Java class files is new and not addressed in classical debuggers.

While no one has investigated the query-based debugging specifically, various researchers have proposed a variety of enhancements to conventional debugging [And95, DHKV93, GH93, GWM89, KRR94, Laf97, LM94, LN97]. The debuggers most closely related to dynamic query-based debugging visualize object relationships—usually references or an object call graph. Duel [GH93] builds on gdb facilities to display data structures by using user script code at a breakpoint. HotWire [LM94] allows users to specify custom object visualizations in constraint language. Look! [And95] supports adding breakpoints, filters and watch windows at runtime for C++ debugging. Object Visualizer [DHKV93], PV [KRR94], and Program Explorer [LN97] provide numerous graphical and statistical run-time views with class-dependent filtering but do not allow general queries. Our debugger can gather statistical data through queries with non-empty results ("How many lists of size greater than 500 exist in the program?") but does not display animated statistical views.

Debuggers that gather information by either instrumenting the source code [DHKV93, LM94] or by using program traces [KRR94, LN97] usually require program recompilation for each different view and do not allow on-the-fly modifications. Gamma et. al. [GWM89] allow different on-the-fly views of debugged programs based on ET++ framework. Laffra [Laf97] discusses visual debugging in Java using source code instrumentation or JVM changes. As mentioned above, JVM modifications would support on-the-fly debugging, but would make the tool dependent on the modified JVM. We opted for the portable method—class file instrumentation at load time. The instrumentation is done by providing a custom class loader. Other non-portable load-time instrumentation alternatives were comprehensively explored by Duncan and Hölzle [DH99].

Consens et al. [CHM94, CMR92] use the $Hy^+$ visualization system to find errors using post-mortem event traces. De Pauw et al. [DLVW98] and Walker et al. [WM+98] use program event traces to visualize program execution patterns and event-based object relationships, such as method invocations and object creation. Similarly, Bruegge and Hibbard [BH83] use generalized path expressions that refer to program events and variables to check program execution path correctness. The path expressions are powerful tool allowing to identify program statements and variables in different procedure invocations. Path rules can be enabled on-the-fly. However, path expressions do not allow optimized join query evaluations or selections on large groups of objects. The work on trace analysis is complementary to ours because it focuses on querying and visualizing run-time events while we query object relationships.

Pre-/postconditions and class invariants as provided in Eiffel [Mey88] can be thought of as language-supported dynamic queries that are checked at the beginning or end of methods. Unlike dynamic queries, they are not continuously checked, they cannot access objects unreachable by references from the checked class, nor can they invoke arbitrary methods. Assertions and invariants cannot be invoked on-the-fly. Dynamic queries could be used to implement class assertions for languages that do not provide them and to support assertions active only during part of the execution.

Dynamic queries are related to incremental join result recalculation in databases [BC79, BLT86]. We use the basic insights of this work to implement the incremental query evaluation scheme. Database queries and views automatically support "on-the-fly" functionality.

# 7 Conclusions

The cause-effect gap between the time when a program error occurs and the time when it becomes apparent to the programmer makes many program errors hard to find. The situation is further complicated by the increasing use of large class libraries and complicated pointer-linked data structures in modern object-oriented systems. A misdirected reference that violates an abstract relationship between objects may remain undiscovered until much later in the program's execution. Conventional debugging methods offer only limited help in finding such errors. Our previous work described a dynamic query-based debugger that allows programmers to ask queries about the program state and updates query results whenever the program changes an object relevant to the query, helping programmers to discover object relationship failures as soon as they happen. The debugger also helps users to watch the changes in object configurations through the program's lifetime. This functionality can be used to better understand program behavior.

The implementation of the query-based debugger is portable across Java virtual machines and operating systems and has good performance. Selection queries are efficient with less than a factor of two slowdown for most queries measured. Join queries are practical when domain sizes are small and queried field changes are infrequent.

This paper extends the debugger to allow programmers to ask queries on-the-fly. This achieves the following goals:

- Interactivity - allows programmers to ask queries in the middle of the program execution to exclude runtime periods when the query is not satisfied or to increase the query's efficiency by enabling it only in a small part of the execution.

- Portability - provides on-the-fly functionality in a portable way for different operating systems and Java virtual machines at a reasonable cost.

These features of the on-the-fly debugger makes it attractive for object-oriented debugging needs.

On-the-fly debugging still has a relatively high overhead. The instrumented programs with inactive debugger suffer a median overhead of 25% (less than 70% for all applications). An enabled debugger that never evaluates a query (for example, because the query references only non-instantiated classes) has a median overhead of 62% with maximum of factor 3 overhead. Selection slowdowns range up to factor 9.5 with a median of 5.5. The on-the-fly debugger has four times higher overhead than the dynamic query-based debugger. Further optimizations could reduce this overhead. The tool is practical for short program runs and infrequently evaluated queries.

We believe that dynamic query-based debugging adds another powerful tool to the programmer's tool chest for tackling the complex task of debugging. We hope that future mainstream debuggers will integrate a similar functionality, simplifying the difficult task of debugging and facilitating the development of more robust object-oriented systems.

## 8   Acknowledgments

We thank Karel Driesen for valuable comments on this paper. This work was funded in part by Nokia Research Center, Sun Microsystems, the State of California MICRO program, and by the National Science Foundation under CAREER grant CCR96–24458 and grants CCR92–21657 and CCR95–05807.